\documentclass[prd,preprint,aps,dvips]{revtex4}
\usepackage[english]{babel}
\usepackage{amscd}
\usepackage{epsfig}
\usepackage{tabularx}

\usepackage{graphicx}
\usepackage{latexsym}
\usepackage{amsmath}
\usepackage{amsfonts}
\usepackage{amssymb}
\usepackage[latin1]{inputenc}

\def\B{\vec B}
\def\E{\vec E}
\def\grad{\vec\nabla}

\newcommand{\be}{\begin{equation}}
\newcommand{\ee}{ \end{equation}}
\newcommand{\ben}{\begin{eqnarray}}
\newcommand{\een}{\end{eqnarray}}
\newcommand{\sech}{\rm sech}
\newcommand{\arcsinh}{\rm arcsinh}
\newcommand{\arctanh}{\rm arctanh}

\begin{document}

\title{Deforming tachyon kinks and tachyon potentials}
\author{V.I. Afonso$^{a}$, D. Bazeia$^{a}$ and F.A. Brito$^{b}$}

\affiliation{$^a$Departamento de F\'\i sica, Universidade Federal
da Para\'\i ba, Caixa Postal 5008, 58051-970 Jo\~ao Pessoa,
Para\'\i ba, Brazil
\\
$^b${Departamento de F\'\i sica, Universidade Federal de Campina
Grande,} Caixa Postal 10071, {58109-970 Campina Grande, Para\'\i ba,
Brazil}}

\date{\today}

\begin{abstract}
In this paper we investigate deformation of tachyon potentials and
tachyon kink solutions. We consider the deformation of a DBI type
action with gauge and tachyon fields living on D1-brane and D3-brane
world-volume. We deform tachyon potentials to get other consistent
tachyon potentials by using properly a deformation function
depending on the gauge field components. Resolutions of singular
tachyon kinks via deformation and applications of deformed tachyon
potentials to scalar cosmology scenario are discussed.
\end{abstract}


\maketitle

\section{Introduction}

It has been conceived in the literature \cite{sen1,sen_papers} that
the D-branes in string theory can be viewed as realizations of
tachyon kinks \cite{kim0,tach1,kim,tach2,tach3,tach4,tach5,tach6}.
Higher dimensional branes can be related to lower dimensional ones
via descent relations \cite{sen}. Unstable D$p$-branes may allow for
stable or unstable D$(p-1)$-brane solutions on their world-volume.
This can be understood, as for example, a picture where stable type
IIB D$(p-1)$-branes with ``right dimension'' (odd dimension) may
live on a unstable type IIB D$p$-brane with ``wrong dimension''
(even dimension). Under this consideration, in order to establish
the equivalence of tachyon kinks and D$(p-1)$-branes, it is
necessary to find tachyon kink solutions with finite energy per unit
$(p-1)$-volume \cite{sen}. Thus, it is interesting to investigate
the properties of tachyon kink solutions in some tachyon models. In
particular, it would be interesting to find a way of `resolving' a
singular tachyon kink solution whose energy is infinite via
deformation of tachyon potentials. As we shall show later,  we can
introduce a deformation function associated with electromagnetic
fields that can play this role. Thus, a singular tachyon kink of a
theory describing a D$p$-brane can be smoothed out by properly
deforming the tachyon potential, living on the D$p$-brane
world-volume, into a new potential of a deformed theory describing a
``deformed D$p$-brane''. We show that while a deformed and smooth
tachyon kink can confine electromagnetic fields on its world-volume,
the singular tachyon kink of the original (non-deformed) theory
cannot. Specially, for a D3-brane the electromagnetic field
components can be localized everywhere on its world-volume. However,
as we deform the tachyon potential, the world-volume of the deformed
D3-brane may not localize all the components of the gauge field.
This is because the smooth tachyon kink that we identify with a
D2-brane living  on the world-volume of the deformed D3-brane, now
can trap components of the gauge field. For instance, for some
choice of parameters one might have a D2-brane confining a magnetic
component or an electric component of the gauge field. This would
characterize an alternative to compactification of a deformed
D3-brane into a deformed D2-brane (the smooth tachyon kink) as a
magnetically or an electrically charged D2-brane. A similar thing
happens as one wraps strings living on the world-volume of a
D$p$-brane carrying a gauge field. A string will become magnetically
or electrically charged depending on the direction of the D$p$-brane
world-volume manifold the string is wrapped.

In this paper we investigate deformed tachyon potentials and
deformed tachyon kinks, following the lines of developments on
tachyon kinks put forward in \cite{kim0,kim}. Our main goal here is
to address the aforementioned issue just by deforming ``string
theory motivated'' tachyon potentials to get other consistent
tachyon potentials by using properly the functional form of the
gauge field components. We also discuss possible applications of
deformed tachyon potentials to scalar cosmology scenario. The paper
is organized as follows. In Sec.~\ref{prel} we consider the
energy-momentum tensor and equations of motion for a DBI type action
with gauge and tachyon fields living in a two-dimensional
world-volume of a D1-brane. In Sec.~\ref{deftachpot} we develop a
general formalism to deform tachyon potentials by making use of a
deformation function. The Sec.~\ref{csimons} is devoted to a
generalization of the DBI action by adding a Chern-Simons like term,
which makes necessary to perform deformation of tachyon
kinks/potentials with non-constant gauge field components. In
Sec.~\ref{sec.def_tach} we look for a tachyon kink solution and its
deformed counterpart. The Sec.~\ref{sec.ext_hig} generalizes the
treatment of Sec.~\ref{deftachpot} for higher dimensional branes,
specially D3-branes. In Sec.~\ref{concu} we make our final comments,
considering some issues in tachyon cosmology.

\section{Preliminaries}
\label{prel}

Let us consider a DBI type action describing tachyons and gauge
fields on the world-volume of a non-BPS D$p$-brane
\cite{garousi2000,bergs} \ben\label{S}
S=-T_p\int{d^{p+1}x\,V(T)\sqrt{-\det{(\eta_{\mu\nu}+\partial_\mu
T\partial_\nu T + F_{\mu\nu}})}}. \een For electric and tachyon
static fields living on the world-volume of a D$1$-brane (or
D-string) we have the Lagrangian \ben\label{Sx} {\cal
L}=-T_1\,V(T)\sqrt{1+{T'}^2 -F_{01}^2}, \een where $V(T)$ is the
tachyon potential, $T\equiv T(x)$ and $E\equiv E_1(x)\!=\!F_{01}$.
The equations of motion are \ben \label{eomT} \left(\frac{V(T)
T'}{\sqrt{1+{T' }^2-E^2}}\right)'-{\,V'(T)}\sqrt{1+{T'}^2-E^2}=0,
\\
\label{eomE} \Pi'=\left(T_1\frac{V(T) E
}{\sqrt{1+{T'}^2-E^2}}\right)'=0, \een where the prime means
derivative with respect to the argument of the function. The
conjugate momentum $\Pi$ is constant. The non-vanishing
energy-momentum tensor components are \ben \label{T00}
T_{00}=T_1\frac{V(T)(1+{T'}^2)}{\sqrt{1+{T'}^2-E^2}},\\
\label{T11} T_{11}=-T_1\frac{V(T)}{\sqrt{1+{T'}^2-E^2}}. \een Notice
that from equations (\ref{eomE}) and (\ref{T11}) and from the
conservation of the energy-momentum tensor $\partial_\mu
T^{\mu\nu}\!=\!0$ we find the following constraint \ben\label{T11p}
\label{cst} -T_{11}'=\left(\frac{\Pi}{E}\right)'=0. \een This
constraint together with (\ref{eomE}) forces the electric field $E$
to be constant, unless we include some ``external source'' into
equation (\ref{eomE}). Indeed, this is achieved via Chern-Simons
like terms, that we shall consider later, since we plan to focus on
non-constant electric and magnetic fields.

\section{Deformed tachyon potentials}
\label{deftachpot}

In this section we address the issue of transforming the DBI type
action for tachyons and gauge fields into another DBI type action
with only a deformed tachyon field. In the ``deformed theory'' the
gauge fields are encoded into a ``deformation function''. We can
transform the Lagrangian (\ref{Sx}) into a deformed one as follows.
Firstly we rewrite (\ref{Sx}) as \ben\label{Sd} {\cal
L}&=&-T_1\,V(T)\sqrt{1-E^2}\sqrt{1+\frac{{T'}^2}{
1-E^2}}\nonumber\\
&=&-T_1\,V(\widetilde{T})\sqrt{1-E^2}\sqrt{1+{\widetilde{T'}}^2}.\een
Let us now consider the transformation \ben\label{VT}{\cal
L}\to\widetilde{\cal L},\qquad V(\widetilde{T})\sqrt{1-E^2}\to
\,\widetilde{V}(\widetilde{T})=\frac{V(T)}{\lambda \sqrt{1-E^2}}.
\een This transformation leads to a new theory given by the
Lagrangian \ben\label{Sxtil} \widetilde{\cal
L}=-T_1{\,\widetilde{V}(\widetilde{T})\sqrt{1+{\widetilde{T'}}^2}}.
\een We regard this theory as ``deformed theory'', with the deformed
tachyon field \ben \label{Ttil}
\widetilde{T}=\pm\int{\frac{dT}{\sqrt{1-E^2}}},\een
 and deformed tachyon potential $\widetilde{V}(\widetilde{T})$.
The transformation (\ref{VT}) is justified by the fact we are
requiring that both theories (\ref{Sd}) and (\ref{Sxtil}) maintain
their energy-momentum components $T_{11}$ and $\widetilde{T}_{11}$
conserved. The pressures $T_{11}$ and $\widetilde{T}_{11}$ are
constants related to each other via real parameter $\lambda>0$. We
can check this explicitly, i.e.,
\ben\label{T11til}\widetilde{T}_{11}=-T_1\frac{\widetilde{V}(\widetilde{T})}
{\sqrt{1+\widetilde{T'}^2}}=-T_1\frac{V(T)}{\lambda\sqrt{1-E^2}}
\frac{1}{\sqrt{1+\frac{{T'}^2}{1-E^2}}}=-T_1\frac{V(T)}{\lambda\sqrt{1-E^2+{T'}^2}}
=\frac{T_{11}}{\lambda}. \een  In this work we are interested in
resolving tachyon solutions with nonvanishing pressure $T_{11}$,
which show divergent energy density --- the pressureless case is
treated by Sen in Ref.~\cite{sen03}. Thus the DBI type action with
electric and tachyon fields can be transformed to another with only
a deformed tachyon field. We summarize the transformations above as
follows: \ben\label{transfT}
d\widetilde{T}=\pm\frac{dT}{\sqrt{1-E^2}}=\pm\frac{\lambda dT}
{[f'(\widetilde{T})]^2},\\
\label{transfV}\widetilde{V}(\widetilde{T})=\frac{V(T)}{\lambda\sqrt{1-E^2}}=\frac{V(T)}
{[f'(\widetilde{T})]^2}, \een where $\widetilde{T}\!=\!f^{-1}(T)$.
Here we have followed the idea of  deformed defects put forward in a
former investigation \cite{dd} --- see also \cite{dd1,dd2}. We
regard the function $f(\widetilde{T})$ as the deformation function
defined as \ben\label{fd}
f(\widetilde{T})=\int{\lambda^{1/2}(1-E^2)^{1/4}}d\widetilde{T}.\een
If $E$ is constant this means nothing but
$f(\widetilde{T})\!=\!\lambda^{1/2}{(1-E^2)}^{1/4}\widetilde{T}$ .
However, as we can see below, there are interesting deformations if
we consider non-constant gauge fields.

An important feature we want to emphasize here is that we are
deforming a singular kink with infinite energy to a smooth kink with
finite energy. Note that this can be achieved because the energy
densities of the deformed and non-deformed theories are given by
\ben\label{T00t}\widetilde{T}_{00}=-\widetilde{T}_{11}(1+{\widetilde{T'}}^2)
=-\frac{T_{11}}{\lambda}\left(1+\frac{{T'}^2}{{f'(\widetilde{T})}^2}\right),\qquad
{T}_{00}=-{T}_{11}(1+{T'}^2), \een where we have used the equations
(\ref{T00}), (\ref{T11}), (\ref{T11til}) and the fact that the
deformation function defines the map $\widetilde{T}\!=\!f^{-1}(T)$.
Supposing that $\widetilde{T}$ is a smooth tachyon kink, such that
behaves asymptotically as
$\widetilde{T'}(|x|\to\widetilde{x}_{vac})=0$, its vacuum energy
density $\widetilde{T}_{00}=-\widetilde{T}_{11}$ is clearly finite.
On the other hand, if $T$ is a singular tachyon kink behaving as
${T\,'}(|x|\to\widetilde{x}_{vac})\to\infty$, with nonvanishing
pressure, then $T_{00}\to\infty$. These statements can only be
consistent with (\ref{T00t}) if
\ben\label{limit}\lim_{|x|\to\widetilde{x}_{vac}}\frac{{T'}^2}{{f'(\widetilde{T})^2}}=0.\een
This is always true for the simple case $f'(\widetilde{T})\propto
{(T')}^{\,\gamma}$,  with $1<\gamma<2$. For
$\widetilde{T'}(|x|\to\widetilde{x}_{vac})\propto
-|x|+\widetilde{x}_{vac}$, we find the simple deformation function
$f(x)\propto
{x}\,({-|x|+\widetilde{x}_{vac}})^{\frac{2-\gamma}{1-\gamma}}$.

\section{Brane within branes and Chern-Simons like couplings}
\label{csimons}

There exists a remarkable feature in D-branes systems, that is the
possibility of constructing D-branes with lower or higher dimension
from others with a fixed dimension (see Refs.~\cite{cjohnson} for
comprehensive reviews, and references therein). This is possible by
adding Chern-Simon like terms \cite{green_hull} to the action
(\ref{S}), which are couplings of the D$p\,$-brane to background
Ramond-Ramond (R--R) fields, e.g., \ben\label{S_WZ}
S=-T_p\int{d^{p+1}x\,V(T)\sqrt{-\det{(\eta_{\mu\nu}+\partial_\mu
T\partial_\nu T + F_{\mu\nu}})}}+T_p\int{C_{(p-1)}}\wedge F, \een
where $F=dA_{(1)}$ is the Abelian Born-Infeld 2-form field strength
on the D$p$-brane and $C_{(p-1)}$ is a R--R $(p-1)$-form. The R--R
potential $C_{p-1}$ acts as a source of a ``dissolved''
D($p-2$)-brane living in the world-volume of a D$p$-brane. The
Chern-Simons like part of the action above is indeed a BF model
\cite{blau,baez} which is topological, i.e., a metric independent
action $\int_M B_q\wedge dA_{n-q-1}$ on a $n$-dimensional manifold
$M$. Here, the manifold $M$ is our $(p+1)$-dimensional D$p$-brane
world-volume. Since the energy-momentum tensor is given by
$(1/\sqrt{-g}\,)\,\delta S/\delta g^{\mu\nu}$, where
$g_{\mu\nu}=\eta_{\mu\nu}+\partial_\mu T\partial_\nu T$, and the
topological BF term is metric independent, the components
(\ref{T00}) and (\ref{T11}) remain unchanged. Note that the equation
of motion of the tachyon field does not change either, because this
term does not depend on the tachyon field explicitly. For a
D$1$-brane, we simply write down the Lagrangian \ben\label{Sx_WZ}
{\cal L}=-T_1{\,V(T)\sqrt{1+{T'}^2 -F_{01}^2}}
+T_1{C_{0}(x)}F_{01}.\een Now $C_{(0)}$ is a scalar potential from
the R--R sector of a type IIB string theory
\cite{green_hull,cjohnson}. The equation of motion of the electric
field now reads \ben \label{eomE_WZ} \Pi'=\left(T_1\frac{V(T) E
}{\sqrt{1+{T'}^2-E^2}}-T_1C_0(x)\right)'=0.\een The non-constant
background R--R scalar field $C_0(x)$ acts as a source of a
D($-1$)-brane (a Dirichlet instanton in type IIB string theory).
Since now we have $\Pi\!=-T_{11}E-T_1C_0(x)$, the equation
(\ref{eomE_WZ}) does not force the electric field to be constant
anymore; it suffices that $T_{11}E(x)=-T_1C_0(x)$ to hold
(\ref{eomE_WZ}). This equation relates the electric field on the
D1-brane with dissolved D($-1$)-branes living on the D1-brane, that
agrees with earlier discussion about dissolved branes.

\section{Deformed Tachyon Kinks}
\label{sec.def_tach}

We are now ready to look for deformed tachyon kinks for non-constant
electric field. This extends the analysis developed in
\cite{kim0,kim}. Let us first rewrite the equation (\ref{T11}) as
\ben \label{Etot} \varepsilon=\frac{1}{2}{T'}^2+U(E,T). \een This is
similar to the total energy of a particle of mass equal to unity,
with $\varepsilon\!=\!-1/2$ and potential energy \ben\label{UE}
U(E,T)=-\frac{1}{2}\left[\frac{T_1 V(T)}{-T_{11}}
\right]^2-\frac{1}{2}E^2.\een From the total energy (\ref{Etot}) we
can reduce the problem of finding tachyon kinks solutions to the
problem of solving a first-order differential equation consistent
with the tachyon equation (\ref{eomT}). The solutions are found via
equation \ben \label{eom1st}
\int_0^T{\frac{d{T}}{\sqrt{\left[\frac{T_1V(T)}{-T_{11}}\right]^2-(1-E^2)}}}=\pm\,
x. \een Now using the transformations (\ref{transfT}) and
(\ref{transfV}) we can easily rewrite (\ref{eom1st}) as \ben
\label{eom2nd}\int_0^{\widetilde{T}}{\frac{d\widetilde{T}}
{\sqrt{\left[\frac{\lambda T_1\widetilde{V}(\widetilde{T})}
{-T_{11}}\right]^2-1}}}=\pm\, x.\een Notice that this is equivalent
to work with the equation (\ref{T11til}) for the
$\widetilde{T}_{11}$ ($=T_{11}/\lambda$). By using the equation
(\ref{eom2nd}), up to difficulties with integrability, we should be
able to find deformed tachyon kinks $\widetilde{T}$ from deformed
tachyon potentials $\widetilde{V}$.

Let us now consider some examples. First we consider the following
tachyon potential in the DBI type action given in (\ref{Sx_WZ})
\ben\label{potEx1} V(T)={\sech}({T/T_0}).\een The deformed potential
is \ben\label{Vex1} \widetilde{V}(\widetilde{T})=\frac{
{\sech}({T/T_0})}{[f'(\widetilde{ T})]^2}=\frac{
{\sech}({T/T_0})}{\sqrt{1-E^2}}.\een

If the electric field $E$ is
constant and $\lambda\!=\!1$ as considered in \cite{kim0,kim}, the
equation (\ref{eom2nd}) turns out to be \ben\label{eom2ndEx1Econs}
\int_0^T{\frac{d{T}}{\sqrt{\frac{E^2T_1^2}
{{\Pi}^2({1-E^2})}\,{\sech}^2({T/T_0})-1}}}=\pm\,\sqrt{1-E^2}x,\een
where we have used $T_{11}\!=\!\Pi/E$. Working out the equation
(\ref{eom2ndEx1Econs}) gives \ben \label{solEx1}
T(x)=T_0\,{\arcsinh}{\left[\sqrt{\frac{E^2T_1^2}{\Pi^2(1-E^2)}-1}\,
\sin{\left(\frac{\sqrt{1-E^2}}{T_0}x\right)} \right]},\een which is
the result found in \cite{kim0,kim}.

\begin{figure}[ht]
\includegraphics[{angle=90,height=5.0cm,angle=180,width=6.0cm}]{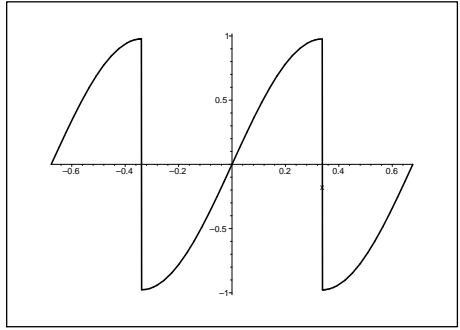}
\caption{The deformed tachyon kink behavior in the interval
$-\pi/\omega\leq x\leq\pi/\omega$, with $T_0\!=\!1$, $T_1\!=\!0.1$,
$T_{11}\!=\!10$ and $\lambda=464$.}\label{fig1}
\end{figure}

Now consider the electric field $E$ is not constant and can be
related to a deformation function according to (\ref{fd}). For a
more general analysis let us assume $V(T)\!=\!{\sech}^q(T/T_0)$. We
choose, among the ones satisfying the criterion (\ref{limit}), the
following deformation function: \ben\label{fdEx1}
T=f(\widetilde{T})=T_0\,{\arctanh}({\widetilde{T}/T_0}).\een As we
will see below, this function allows us to construct analytically a
periodic tachyon kink that is consistent with a tachyon kink wrapped
on a circle. Substituting this into equation (\ref{Vex1}) we find
the deformed tachyon potential
\ben\label{Vex1ET}\widetilde{V}(\widetilde{T})=\frac{
{\sech}^q({T/T_0})}{[f'(\widetilde{T})]^2}={\left(\sqrt{1-\tanh^2(T/T_0)}\right)^q}
\left(1-\frac{\widetilde{T}^2}{T_0^2}\right)^2
=\left(1-\frac{\widetilde{T}^2}{T_0^2}\right)^{\frac{q}{2}+2},\een
which has global minima at $\widetilde{T}\!=\!\pm T_0$ and global
maximum at $\widetilde{T}\!=\!0$, for $\frac{q}{2}+2\!>\!0$. Notice
that this potential is defined only in the interval
$(-{T}_0,{T}_0)$. Substituting this potential into (\ref{eom2nd}) we
find \ben
\label{eom2ndEx1}\int_0^{\widetilde{T}}{\frac{d\widetilde{T}}
{\sqrt{\frac{\lambda^2T_1^2}
{T_{11}^2}\left(1-\frac{\widetilde{T}^2}{T_0^2}\right)^{q+4}-1}}}=\pm\,
x.\een The general solution involves polynomial terms and elliptical
functions, and is not necessary for our discussions here. The case
$q\!=\!1$ which recover the potential (\ref{potEx1}) is still hard
to work with. Let us turn, however, to the simplest case, i.e.,
$q\!=\!-3$ to get the simple nontrivial solution \ben
\label{kink1}\widetilde{T}(x)=
T_0\sqrt{1-\frac{T_{11}^2}{\lambda^2T_1^2}}\frac{\tan(\omega
x)}{\sqrt{\tan(\omega x)^2+1}},\een where
$\omega\!=\!\sqrt{\lambda^2 T_1^2/T_0^2T_{11}^2}$. This is a
periodic tachyon kink whose period is $\pi/\omega$ --- see
Fig.\ref{fig1}, that is in accord with a tachyon kink on a circle of
radius $R=1/2\omega$ \cite{sen}. This tachyon kink is stable because
the vacuum manifold of the deformed potential $\widetilde{V}$
consists of a pair of points $\pm{T}_0$, and the solution connects
different vacua. This solution can be recognized as a stable
D0-brane living on the D1-brane (or D-string) world-volume
\cite{sen}. Since the D1-brane is unstable, it may decay leaving
behind a stable D0-brane. This relation between the D1-brane and the
D0-brane is an example of `descent relations' that exist among
D-branes, as first pointed out by Sen \cite{sen}.

\begin{figure}
\includegraphics[{angle=90,height=5.0cm,angle=180,width=6.0cm}]{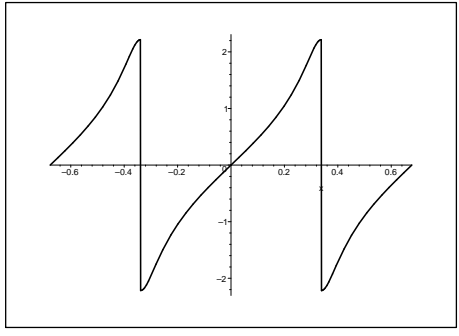}
\caption{The singular tachyon kink behavior in the interval
$-\pi/\omega\leq x\leq\pi/\omega$, with $T_0\!=\!1$, $T_1\!=\!0.1$,
$T_{11}\!=\!10$ and $\lambda=464$.}\label{fig3}
\end{figure}

Using the deformation function (\ref{fdEx1}) one can also obtain the
tachyon kink $T$ of the original theory (\ref{Sx_WZ}) given by
\ben\label{Tach}
T(x)=T_0{\arctanh}\left({\sqrt{1-\frac{T_{11}^2}{\lambda^2T_1^2}}
\frac{\tan(\omega x)}{\sqrt{\tan(\omega x)^2+1}}}\right). \een This
tachyon kink becomes singular for
${T_{11}^2}/{\lambda^2T_1^2}=1/\omega^2T_0^2=4R^2/T_0^2$
sufficiently small, i.e., $R\ll T_0/2$
--- see Fig.~\ref{fig3}. It is singular in the sense that its energy
diverges, whereas the deformed tachyon kink is smooth and has finite
energy.  A singular tachyon kink wrapped on a circle $S^1$ with
radius $R\ll T_0/2$ of a D$(p+1)$-brane world-volume manifold
$S^1\times \mathbb{M}^{p+1}$ becomes ``resolved'' in the deformed
theory with the same compactification radius. Thus we can think of
the deformation process as a way of ``smoothing out'' singular
tachyon kink solutions through gauge fields. This is similar to
``brane resolution'', where singularity of branes can be resolved by
turning on fluxes \cite{cvetic_poper_etal}.

\begin{figure}
\includegraphics[{angle=90,height=5.0cm,angle=180,width=6.0cm}]{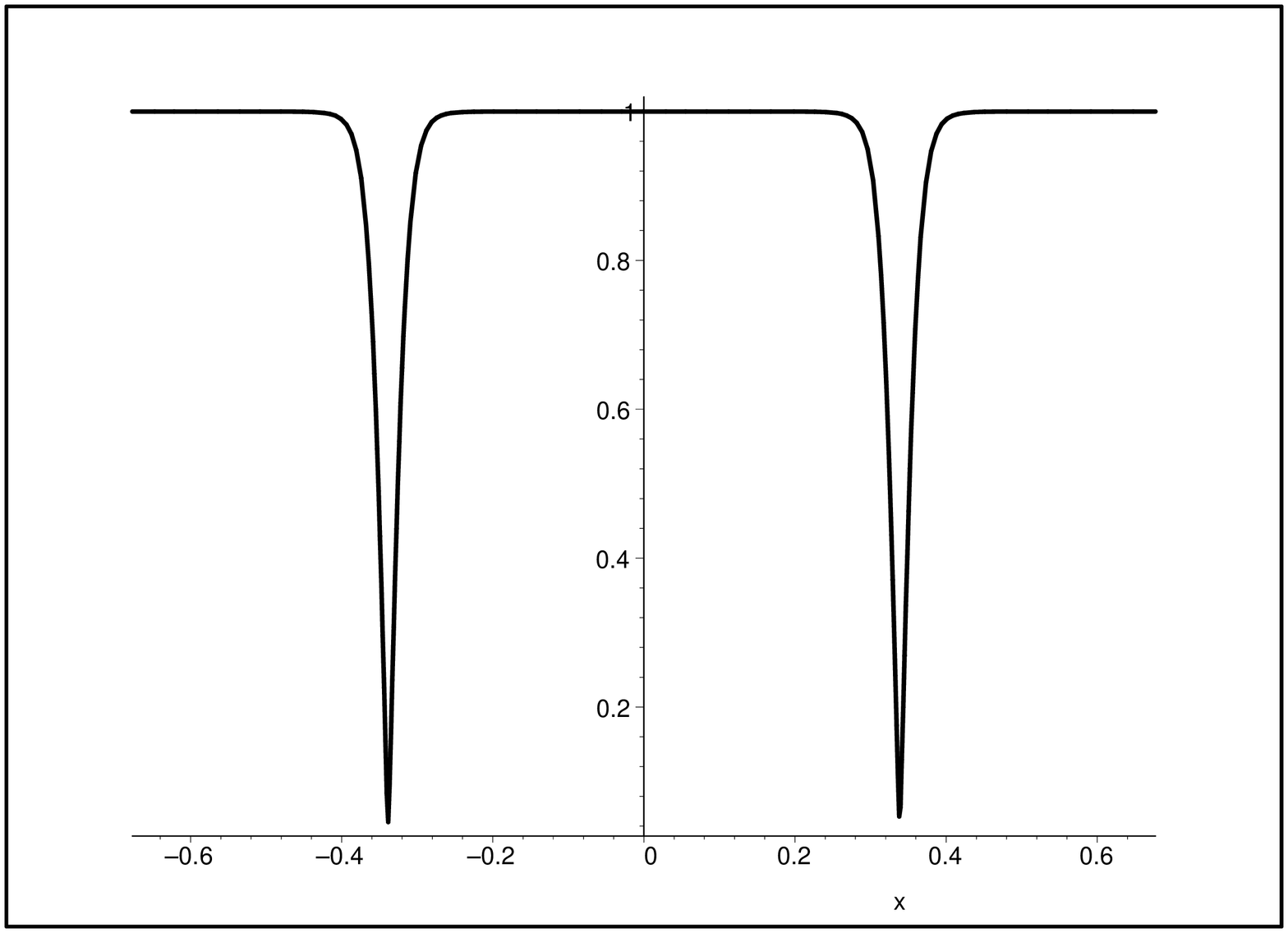}
\caption{The electric field on the deformed tachyon kink, in the
interval $-\pi/\omega\leq x\leq\pi/\omega$, for $T_0\!=\!1$,
$T_1\!=\!0.1$, $T_{11}\!=\!10$ and $\lambda=464$.}\label{fig2}
\end{figure}

The electric field can be calculated via Eq.(\ref{fd}). Applying the
deformation function $f(\widetilde{T})$ defined as in (\ref{fdEx1})
gives \ben\label{Efield} E=
\sqrt{1-\lambda^{-2}\left(1-\frac{{\widetilde{T}}^2}{T_0^2}\right)^{-4}}.
\een Substituting the deformed tachyon kink solution (\ref{kink1})
into (\ref{Efield}), the electric field $E(x)$ develops the behavior
depicted in Fig.~\ref{fig2}.
The electric field is confined to the new lower dimensional D0-brane
--- note it falls off fast along the transverse coordinate $x$. Both
kink solution (\ref{kink1}) and electric field function
(\ref{Efield}) impose constraints on the parameter $\lambda$, i.e.,
\ben\label{lambda} \frac{T_{11}}{T_{1}} \leq \lambda \leq
\left(\frac{T_{11}}{T_{1}}\right)^{4/3}, \qquad T_{11}>T_1. \een The
electric field $E(x)$ we found above is localized either on the
original D3-brane or on the D2-brane that appears as a deformed
tachyon kink living on the D3-brane of the deformed theory. Note
this is consistent with the fact that in the original D3-brane the
tachyon kink is singular and cannot localize (or expel) the electric
field. As a consequence, the electric field is present everywhere on
the D3-brane world-volume and appears explicitly on the DBI action.
On the other hand, the D3-brane world-volume of the the deformed
theory has no explicit dependence on the electric field. This field
now reappears as a localized field on the deformed tachyon kink (the
D2-brane) and cannot be present everywhere on the $deformed$
D3-brane world-volume anymore.


\section{Extension to higher dimensional D-branes}
\label{sec.ext_hig}

Our previous analysis can be extended to D$p$-branes with $p+1$
dimensional world-volume. The Lagrangian for DBI and Chern-Simons
like terms now reads \be \label{LHigher}{\cal L}=-{T_p
V(T)}\sqrt{-X}+\frac{1}{2\,(p-1)!}\epsilon^{\mu_1..\mu_{p+1}}C_{\mu_1..\mu_{p-1}}F_{\mu_{p}\,
\mu_{p+1}}, \ee where in terms of components we write \be \label{X}
-X \equiv 1+(\grad T)^2+(\grad T\cdot\B)^2-(\grad
T\times\E)^2+\B^2-\E^2-(\E\cdot\B)^2. \ee As a concrete example, let
us focus on static field configurations on a D3-brane. The field
equation for the tachyon living on the world-volume of a D3-brane is
\ben \grad\cdot\left[\frac{V(T)}{\sqrt{-X}} \left((1-\E^2)\grad T +
(\E\cdot\grad T) \E + (\B\cdot\grad T) \B \right)\right] - V'(T)
\sqrt{-X}=0. \een
 The conjugate momenta for the gauge fields, $\Pi_i$, can be written as
 \ben \label{Pconj3}
\Pi_1&=&\frac{T_3 V(T)}{\sqrt{-X}}\left[ (1+(\grad T)^2) E_1 + (\E\cdot\B) B_1- (\E\cdot\grad T) (\grad T)_1 \right]+\Pi_{1CS}, \\
\Pi_2&=&\frac{T_3 V(T)}{\sqrt{-X}}\left[ (\E\cdot\B) E_2 -(\B\cdot\grad T) (\grad T)_2- B_2 \right] +\Pi_{2CS},  \\
\Pi_3&=&\frac{T_3 V(T)}{\sqrt{-X}}\left[ (\E\cdot\B) E_3 - (\B\cdot\grad T) (\grad T)_3- B3 \right] +\Pi_{3CS},  \\
\Pi_4&=&\frac{T_3 V(T)}{\sqrt{-X}}\left[ (1+(\grad T)^2) E_3 + (\E\cdot\B) B_3- (\E\cdot\grad T) (\grad T)_3 \right] +\Pi_{4CS}, \\
\Pi_5&=&\frac{T_3 V(T)}{\sqrt{-X}}\left[ (\E\cdot\B) E_1 - (\B\cdot\grad T) (\grad T)_1- B_1 \right]+\Pi_{5CS}, \\
\Pi_6&=&\frac{T_3 V(T)}{\sqrt{-X}}\left[ (1+(\grad T)^2) E_2 + (\E\cdot\B) B_2- (\E\cdot\grad T) (\grad T)_2 \right] +\Pi_{6CS},
\een  where
$\Pi_{iCS}$ are contributions from the Chern-Simons like term.
The components of the energy-momentum tensor turn out to be \ben
T_{00}\equiv\rho&=&\frac{T_3 V(T)}{\sqrt{-X}}\left[1+(\grad T)^2+\B^2
+(\B\cdot\grad T)^2\right],  \\
T_{0i}\equiv P_i&=&\frac{T_3 V(T)}{\sqrt{-X}}\left[(\B\cdot\grad T)
(\E\times\grad T)_i +(\E\times\B)_i \right], \een and
\begin{multline}
T_{ij}=\frac{T_3 V(T)}{\sqrt{-X}} \biggl[ (\grad T)_i(\grad
T)_j-E_iE_j-B_iB_j+ (\E\times\grad T)_i (\E\times\grad T)_j \\-
\left( 1-\E^2 +(\grad T)^2 \right) \delta_{ij} \biggr].
\end{multline}

Let us look for tachyon kink configurations, where we specialize
ourselves on tachyon and gauge fields depending only on one
coordinate, say $x$. The integrability of the system simplifies for
such configurations. We can now write the following: \be \label{Xx}
-X \equiv 1+{T\,'}^2+{T\,'}^2(
B_1^2-E_2^2-E_3^2)+\B^2-\E^2-(\E\cdot\B)^2, \ee and the equations of
motion for tachyon and gauge fields as \ben
\left[\frac{V(T)}{\sqrt{-X}}\bigl( 1-\E^2 + E_1^2 + B_1^2 \bigr)
T\,' \right]' - V'(T) \sqrt{-X}=0, \een \ben \label{Pi1}
\Pi_1'&=&\left\{\frac{T_3 V(T)}{\sqrt{-X}}\left[E_1 + (\E\cdot\B) B_1\right]+C_{32}(x)\right\}'=0,\\
\label{Pi2}
\Pi_2'&=&\left\{\frac{T_3 V(T)}{\sqrt{-X}}\left[(\E\cdot\B) E_2 - B_2 \right]+C_{20}(x)\right\}'=0,\\
\label{Pi3}
\Pi_3'&=&\left\{\frac{T_3 V(T)}{\sqrt{-X}}\left[ (\E\cdot\B) E_3 - B_3 \right]+C_{03}(x)\right\}'=0.
\een

The system has energy density $\rho$ and carries the linear momentum density
$P_i$ given by \ben
T_{00}\equiv\rho&=&\frac{T_3 V(T)}{\sqrt{-X}}\left[1+   \B^2+ (1 + B_1^2) {T\,'}^2\right],  \\
T_{0i}\equiv P_i&=&\frac{T_3 V(T)}{\sqrt{-X}}\left[B_1 T\,'
(\E\times\grad T)_i +(\E\times\B)_i \right]. \een The other
non-vanishing components of the energy-momentum tensor are
\begin{multline}
T_{ij}=\frac{T_3 V(T)}{\sqrt{-X}} \biggl[ {T\,'}^2
\delta_{i1}\delta_{j1} -E_iE_j-B_iB_j+ (\E\times\grad T)_i
(\E\times\grad T)_j \\- \left( 1-\E^2 +{T\,'}^2 \right)
\delta_{ij} \biggr],
\end{multline}
where $(\E\times\grad T)_i = T\,'(E_3 \delta_{i2} - E_2
\delta_{i3})$. The conservation of energy-momentum tensor, i.e.,
$\partial^\mu T_{\mu\nu}=0$, implies the four conserved quantities
$T_{01}=T_{10}$, $T_{11}$, $T_{12}=T_{21}$, and $T_{13}=T_{31}$. As
in the one dimensional example above, now we shall make use of the
quantity $T_{11}$, as well as the quantities $T_{01}$, $T_{12}$,
$T_{13}$ given explicitly by \ben
\label{T01II}&&T_{01}=-T_3\frac{V(T)}{\sqrt{-X}}(B_2E_3-B_3E_2),\\
\label{T11II}&&T_{11}=-T_3\frac{V(T)}{\sqrt{-X}}(1+B_1^2-E_2^2-E_3^2),\\
\label{T12II}&&T_{12}=-T_3\frac{V(T)}{\sqrt{-X}}(E_1E_2+B_1B_2),\\
\label{T13II}&&T_{13}=-T_3\frac{V(T)}{\sqrt{-X}}(E_1E_3+B_1B_3).
\een These are constants of motion that will be useful for
integrability of the system, as we shall see below. In addition to
these four independent conserved quantities, there are other three
constants coming from the Faraday's law, i.e., $\partial^\mu
F^*_{\mu\nu}=0$. The constants are the gauge field components $E_2$,
$E_3$ and $B_1$, such that we are left with only $E_1(x)$, $B_2(x)$
and $B_3(x)$ as possible ``dynamical'' components. As a consequence
of such constant components, from (\ref{T11II}) we note that
${V(T)}/{\sqrt{-X}}$ is also constant.

\begin{figure}[empty]
\includegraphics[{angle=90,height=5.0cm,angle=180,width=6.0cm}]{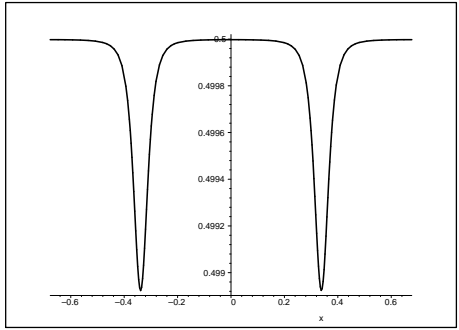}
\caption{The behavior of the electric field component $E_1(x)$
around the deformed tachyon kink (at $x\thickapprox0$), in the
interval $-\pi/\omega\leq x\leq\pi/\omega$, for $E_3\!=\!1$,
$T_0\!=\!1$, $T_3\!=\!0.1$, $T_{13}\!=\!T_{11}\!=\!10$, and
$\lambda\!=\!464$.}\label{fig4}
\end{figure}

\begin{figure}[empty]
\includegraphics[{angle=90,height=5.0cm,angle=180,width=6.0cm}]{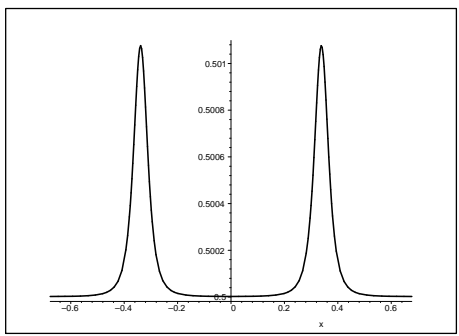}
\caption{The behavior of the magnetic field component $B_3(x)$
around the deformed tachyon kink (at $x\thickapprox0$), in the
interval $-\pi/\omega\leq x\leq\pi/\omega$, for $E_3\!=\!1$,
$T_0\!=\!1$, $T_3\!=\!0.1$, $T_{13}\!=\!T_{11}\!=\!10$, and
$\lambda\!=\!464$.}\label{fig5}
\end{figure}

The generalization of the discussion of Sec.~\ref{deftachpot} to
higher dimensions is straightforward. Specially in three spatial
dimensions, to include new field components, we consider the
following. Let us first rewrite (\ref{Xx}) as
\ben\label{Xx2}-X=\beta+\alpha\, {T'}^2, \een where
$\alpha=1+B_1^2-E_2^2-E_3^2,$ and $\beta=1+\B^2-\E^2
-(\E\cdot\B)^2.$ Thus, for a D3-brane we can write (\ref{LHigher})
as \ben\label{SdII} {\cal
L}&=&-T_3\,V(T)\sqrt{\beta}\sqrt{1+\frac{\alpha{T'}^2}{\beta
}}+{\cal L}_{CS}\nonumber\\
&=&-T_3\,V(\widetilde{T})\sqrt{\beta}\sqrt{1+{\widetilde{T'}}^2}+{\cal
L}_{CS},\een where ${\cal L}_{CS}$ is the Chern-Simons like
Lagrangian. Following earlier steps, we can find a new related
theory by making the deformation \ben\label{VTII}{\cal
L}\to\widetilde{\cal L},\qquad V(\widetilde{T})\sqrt{\beta}\to
\,\widetilde{V}(\widetilde{T})=\frac{\alpha V(T)}{\lambda
\sqrt{\beta}},\een where the deformed theory is given by
\ben\label{SxtilII} \widetilde{\cal
L}=-T_3{\,\widetilde{V}(\widetilde{T})\sqrt{1+{\widetilde{T'}}^2}}+{\cal
L}_{CS}. \een Note that the Chern-Simons like term that couples to
the original D3-brane, is not affected by the deformation. However,
this may not be true for other coupling terms, such as $dT\wedge F$,
involving tachyon field, also used in the literature
\cite{bergs,sen}. The deformed tachyon field is defined as \ben
\label{TtilII}
\widetilde{T}=\pm\int{\sqrt{\frac{\alpha}{\beta}}}\,{dT},\een and
the deformed tachyon potential is $\widetilde{V}(\widetilde{T})$.
Again, as previously considered, by performing the transformation
(or deformation) (\ref{VTII}) both theories (\ref{SdII}) and
(\ref{SxtilII}) maintain their energy-momentum components $T_{11}$
and $\widetilde{T}_{11}$ conserved. The pressures $T_{11}$ and
$\widetilde{T}_{11}$ are constants related to each other via real
parameter $\lambda$ as we can check explicitly:
\ben\label{T11tilII}\widetilde{T}_{11}=-T_3\frac{\widetilde{V}(\widetilde{T})}
{\sqrt{1+\widetilde{T'}^2}}=-T_3\frac{\alpha
V(T)}{\lambda\sqrt{\beta}}
\frac{1}{\sqrt{1+\frac{\alpha{T'}^2}{\beta}}}=-T_3\frac{\alpha
V(T)}{\lambda\sqrt{\beta+\alpha{T'}^2}} =\frac{T_{11}}{\lambda},
\een with $T_{11}$  given in (\ref{T11II}). The transformations
(\ref{transfT})-(\ref{transfV}) generalize and are now given by
\ben\label{transfTII}
d\widetilde{T}=\pm\sqrt{\frac{\alpha}{\beta}}\,dT=
\pm\frac{\lambda\alpha^{-1/2}dT}
{[f'(\widetilde{T})]^2},\\
\label{transfVII}\widetilde{V}(\widetilde{T})=\frac{\alpha
V(T)}{\lambda\sqrt{\beta}} =\frac{V(T)}{[f'(\widetilde{T})]^2}, \een
where $\widetilde{T}\!=\!f^{-1}(T)$, and the deformation function
$f(\widetilde{T})$ is defined as \ben\label{fdII}
f(\widetilde{T})=\int{\lambda^{1/2}\frac{\beta^{1/4}}{\alpha^{1/2}}}
\,d\widetilde{T}.\een Before going into calculations some comments
are in order. Note that since the deformation function captures all
the information related to the gauge fields, the previous
calculations concerning deformed tachyon kinks continues valid here.
The field components contributes to a same deformation. The novelty
is that now we can deal with the effects of such components
individually. The whole deformation is related to the field
components as
\ben\label{EM}[f'(\widetilde{T})]^4\equiv\lambda^2\frac{\beta}{\alpha^2}
=\lambda^2\frac{1+\B^2-\E^2
-(\E\cdot\B)^2}{(1+B_1^2-E_2^2-E_3^2)^2}.\een Note that for $\B=0$,
$E_1=E$ and $E_2=E_3=0$ this equation recovers the equation
(\ref{Efield}). The problem with all field components involve six
variables $(E_1,E_2,E_3,B_1,B_2,B_3)$. So in principle we need six
equations to solve the problem. However, the equation (\ref{EM})
together with the independent conserved quantities we found before
are sufficient to solve the system. Using the fact that $B_1$, $E_2$
and $E_3$ are constants, a simple nontrivial and consistent example
is given by setting $E_2=0$, $B_2=0$ and $B_1=E_3$, that leaves us
with the following relevant equations for $E_1(x)$ and $B_3(x)$ \ben
[f'(\widetilde{T})]^4 &=&\lambda^2[1+B_3^2-E_1^2
-(E_1E_3+E_3B_3)^2],
 \\ T_{13}&=&T_{11}(E_1 E_3+E_3B_3).
\een These equations can be solved for the dynamical field
components \ben\label{E1B3sol}E_1(x)&=&
\frac{1}{2}\frac{-[f'(\widetilde{T})]^2T_{11}^2E_3^2+\lambda^2(T_{11}^2E_3^2
+T_{13}^2-E_3^2T_{13}^2)}{\lambda^2T_{13}T_{11}E_3},\\
B_3(x)&=&\frac{1}{2}\frac{[f'(\widetilde{T})]^2T_{11}^2E_3^2+\lambda^2(-T_{11}^2E_3^2+T_{13}^2
+E_3^2T_{13}^2)}{\lambda^2T_{13}T_{11}E_3}, \een where $\lambda$,
$T_{11}$, $T_{13}$, and $E_3$ are non-vanishing constants. Thus, up
to these constants, the problem is fully determined as long as we
know the deformation function $f(\widetilde{T})$. Let us now use the
deformation function (\ref{fdEx1}) and the deformed tachyon kink
$\widetilde{T}$ given in (\ref{kink1}), to find the field components
--- note that
$[f'(\widetilde{T})]^2=1/[1-(\widetilde{T}^2/T_0^2)]^2$. The
magnitude of the field components, i.e., $|E_1(x)|$ and $|B_3(x)|$
can be peaked around $x=0$ according to the sign of their concavity
\ben\label{mE}&& d^2_x|E_1(x)|_{x=0}={\rm
sgn\,}[T_{11}^2E_3^2(1-\lambda^2)+\lambda^2T_{13}^2(E_3^2-1)]
\left(1-\frac{T_{11}^2}{\lambda^2T_3^2}\right)\zeta,\\
\label{mB}&& d^2_x|B_3(x)|_{x=0}={\rm
sgn\,}[T_{11}^2E_3^2(1-\lambda^2)+\lambda^2T_{13}^2(E_3^2+1)]
\left(1-\frac{T_{11}^2}{\lambda^2T_3^2}\right)\zeta,\een where
$\zeta=\frac{2|T_{11}||E_3|\omega^2}{\lambda^2|T_{13}|}>0$, and
$T_3$ is the 3-brane tension, i.e., the 3d analog of $T_1$ given in
Sec.~\ref{sec.def_tach}. Thus, replacing $T_1\to T_3$ the constraint
on $\lambda$ given in (\ref{lambda}) implies that
$\frac{T_{11}^2}{\lambda^2T_3^2}<1$ (and $T_{11}>T_{3}$,
$\lambda^2>1$), such that the signs of the concavities
(\ref{mE})-(\ref{mB}) are simply determined by the ``sgn''
prefactor. For example, we can easily see that for $E_3=1$ and
$T_{13}=T_{11}$ the concavities are $d^2_x|E_1(x)|_{x=0}<0$ and
$d^2_x|B_3(x)|_{x=0}>0$. Thus, for this choice of parameters the
only field component that is localized around $x=0$ is $E_1(x)$,
whereas the component $B_3(x)$ is not. They are depicted in
Fig.~\ref{fig4} and Fig.~\ref{fig5}, respectively. In the example
shown in Figs.~\ref{fig4} and ~\ref{fig5}, the electric field
component $E_1(x)$ is confined to the deformed tachyon kink (a
D2-brane), while the magnetic field component is not. $B_3(x)$ is
``peaked'' just outside the D2-brane. As another example, one can
also check that for $E_3=1$ and $T_{13}=nT_{11}$, both components
$B_3(x)$ and $E_1(x)$ get localized on the D2-brane, as long as
$n<(1/\lambda)[(\lambda^2-1)/2]^{1/2}$. Finally, for $T_{13}=T_{11}$
and $E_3^2\sim\lambda^2$ we find that both components $B_3(x)$ and
$E_1(x)$ are expelled from the D2-brane. In summary, the fields $\E$
and $\B$ are not present everywhere on the deformed D3-brane,
because the presence of the deformed D2-brane inside, always
confines (or expels) electromagnetic components. This is only
possible in the deformed theory provided its D2-brane has finite
tension. This is assured if the deformation function used satisfies
criterion (\ref{limit}). As the deformed D3-brane decay it may leave
behind stable D2-branes with electromagnetic components localized on
them.

\section{Conclusions}
\label{concu}

In this paper we addressed the issue of deforming tachyon potentials
and tachyon kinks by using a deformation function given in terms of
gauge field components. We have found that a singular tachyon kink
can be resolved via such deformation. The resolved (non-singular)
tachyon kinks can represent D$p$-branes in string theory
\cite{sen1,sen_papers,sen}.

Concerning tachyon cosmology that makes use of time-dependent
homogeneous tachyons $T(t)$ \cite{sen2000}, some brief comments are
in order. In realistic models of scalar cosmology, it is expected
that the small fluctuations of the scalar field  during the
reheating are stable.  This is because there is the possibility of
the fast growth of inhomogeneous scalar field fluctuations after
inflation. In Ref.~\cite{fks} it was pointed out that tachyon
cosmology, for some string theory motivated tachyon potentials,
indeed does develop such instability. The instability of the
inhomogeneous tachyon fluctuations $\delta T(t,\vec{x})=\int
d^{\,3}k\, T_k(t)\,e^{i\vec{k}.\vec{x}}$ around $T(t)$, governed by
the following equation \ben
\frac{\ddot{T}_k}{1-\dot{T}^2}+\frac{2\,\dot{T}\,\ddot{T}}
{(1-\dot{T}^2)^2}\dot{T}_k+[k^2+(\log V)_{,TT}]T_k=0,\een becomes
non-linear very quickly \cite{fks}. However, as has been shown in
\cite{sami_garousi}, this has to do with the `concavity' $(\log
V)_{,TT}$ of the tachyon potential, which means that for $(\log
V)_{,TT}$ positive (negative) one finds stable (unstable)
fluctuations. In Ref.~\cite{sami_garousi} it was considered a scalar
potential which is essentially a `tachyon potential' with reversed
concavity. For example, a tachyon potential
$V(T)\!=\!e^{-(T/T_0)^2}$ with negative concavity $-2/T_0^2$ is
replaced to another
$\widetilde{V}(\widetilde{T})\!=\!e^{(\widetilde{T}/\widetilde{T}_0)^2}$
with positive concavity $2/\widetilde{T}_0^2$, i.e., a massive
inflaton potential. Given that we have investigated deformation of
tachyon potentials in the previous sections, we can make use of this
tool in order to replace a cosmological scenario to another. Thus, a
tachyon potential motivated by string theory could be deformed into
another by a deformation function that is given in terms of gauge
field components associated with background R-R fields.

In fact, in this paper, we have investigated an example in which the
concavity of potentials indeed changes through deformation. This is
the case $p\!=\!-3$, where the potential
$V(T)=1/{\sech}^{3}{(T/T_0)}$ with positive concavity
$(3/T_0^2)\,{\rm sech}^2{(T/T_0)}$ is deformed into
 the potential
$\widetilde{V}(\widetilde{T})=\left(1-\frac{\widetilde{T}^2}{T_0^2}
\right)^{\frac{1}{2}}$ with negative concavity
$-(\widetilde{T}^2+{T}_0^2)/(-{T}_0^2+\widetilde{T}^2)^2$.

In a certain sense the deformed theory is good for tachyon kinks
(inhomogeneous tachyons), but not good for tachyon cosmology
(homogeneous tachyon). The smooth tachyon kink is generated by the
deformed tachyon potential $\widetilde{V}$ which has negative
concavity, whereas the tachyon cosmology is stable for the
``non-deformed'' theory with tachyon potential $V$ which has
positive concavity.

In summary, in this paper we have found an example where the
deformation of a tachyon theory via gauge field components connects
two different limits. On one hand, in the deformed theory we can
find smooth tachyon kinks representing lower dimensional D-branes,
but without stable cosmology, while on the other hand we have stable
cosmology without smooth lower dimensional D-branes in the original
(non-deformed) theory. In particular, note that the same deformation
that makes the smooth tachyon kinks (or deformed D2-branes) to
localize gauge field components, in some way also prevents stable
cosmology on the deformed D3-brane.  This seems to be consistent
with the fact that for potentials with negative concavity (the
deformed potential), the instability of the tachyon cosmology after
inflation takes place and inhomogeneous tachyon fluctuations grow
very fast. In such scenario the instability may start populating the
universe with extended objects such as deformed tachyon kinks or
D2-branes, which are beyond the perturbative spectrum. This issue
and its higher dimensional extensions are to be further explored
elsewhere.

\acknowledgments

We would like to thank CAPES, CLAF/CNPq, CNPq, PADCT/CNPq, and
PRONEX/CNPq/FAPESQ for financial support.


\begin{thebibliography}{99}

\bibitem{sen1}A. Sen, Int. J. Mod. Phys. A {\bf14,} 4061 (1999);
[arXiv:hep-th/9902105] and JHEP {\bf12,} 027 (1999); [arXiv:hep-th/9911116].
\bibitem{sen_papers}A. Sen, ``Non-BPS States and Branes in
String Theory''; [arXiv:hep-th/9904207] and Int. J. Mod. Phys. A
{\bf20}, 5513 (2005); [arXiv:hep-th/0410103].
\bibitem{kim0} C. Kim, Y. Kim and C.O. Lee, JHEP {\bf0305},
020 (2003); [arXiv:hep-th/0304180].
\bibitem{tach1}P. Brax, J. Mourad, and D.A. Steer,
Phys. Lett. B{\bf575}, 115 (2003); [arXiv:hep-th/0304197].
\bibitem{kim} C. Kim, Y. Kim, O.-K. Kwon and C.O. Lee, JHEP {\bf0311},
034 (2003); [arXiv:hep-th/0305092].
\bibitem{tach2} E.J. Copeland, P.M. Saffin, D.A. Steer,
Phys. Rev. D{\bf68}, 065013 (2003);  [arXiv:hep-th/0306294].
\bibitem{tach3}D. Bazeia, R. Menezes and J.G. Ramos,
Mod. Phys. Lett. A{\bf20}, 467 (2005); [arXiv:hep-th/0401195].
\bibitem{tach4}R. Banerjee, Y. Kim  and O.-K. Kwon,
JHEP {\bf0501}, 023 (2005); [arXiv:hep-th/0407229].
\bibitem{tach5}S. Thomas and J. Ward, JHEP {\bf0510}, 098 (2005);
[arXiv:hep-th/0502228].
\bibitem{tach6}C. Kim, Y. Kim, O.-K. Kwon and
H.-U.Yee, ``Tachyon kinks in boundary string field theroy'',
[arXiv:hep-th/0601206].
\bibitem{sen}A. Sen, Int. J. Mod. Phys. A{\bf14}, 4061 (1999);
[arXiv:hep-th/9902105].
\bibitem{garousi2000}M.R. Garousi, Nucl. Phys. B{\bf584}, 284 (2000);
[arXiv:hep-th/0003122].
\bibitem{bergs} E.A. Bergshoeff, M. de Roo, T.C. de Wit, E. Eyras
and S. Panda,  JHEP {\bf0005},  009 (2000);  [arXiv:hep-th/0003221].
\bibitem{sen03}A. Sen, Phys. Rev. D {\bf68}, 066008
(2003); [arXiv:hep-th/0303057].
\bibitem{dd}D. Bazeia, L. Losano, and J.M.C. Malbouisson, Phys. Rev. D {\bf66,}
101701(R) (2002); [arXiv:hep-th/0209027].
\bibitem{dd1}C.A. Almeida, D. Bazeia, L. Losano, and J.M.C. Malbouisson, Phys. Rev. D {\bf69,}
067702 (2004); [arXiv:hep-th/0405238]. D. Bazeia and L. Losano,
Phys. Rev. D {\bf73,} 025016 (2006); [arXiv:hep-th/0511193].
\bibitem{dd2}A. de Souza Dutra, Phys. Lett. B {\bf626,} 249 (2005);
A. de Souza Dutra and A.C. Amaro de Faria, Jr, Phys. Rev. D {\bf72,}
087701 (2005).
\bibitem{cjohnson} J. Polchinski,  ``TASI lectures on
D-branes'', [arXiv:hep-th/9611050]. C.V. Johnson, ``D-brane
primer'', [arXiv:hep-th/0007170]. W. Taylor and B. Zwiebach,
``D-Branes, tachyons, and string field theory'',
[arXiv:hep-th/0311017].
\bibitem{green_hull} M.B. Green, C.M. Hull and P.K. Townsend,
Phys. Lett. B{\bf382}, 65 (1996); [arXiv:hep-th/9604119].
\bibitem{blau} M. Blau and G. Thompson, Ann. Phys. {\bf205}, 130
(1991).
\bibitem{baez} J.C. Baez, Lect. Notes Phys. {\bf543}, 25
(2000); [arXiv:gr-qc/9905087].
\bibitem{cvetic_poper_etal}  M. Cvetic, H. Lu and C.N. Pope,
Nucl. Phys. B{\bf600},  103 (2001); [arXiv:hep-th/0011023].
\bibitem{sen2000}A. Sen, JHEP {\bf0207}, 065 (2002);
[arXiv:hep-th/0203265].
\bibitem{fks} A. Frolov, L. Kofman and A. Starobinsky,
Phys. Lett. B{\bf545},  8 (2002); [arXiv:hep-th/0204187].
\bibitem{sami_garousi} M.R. Garousi, M. Sami, and S. Tsujikawa,
Phys. Rev. D{\bf70}, 043536 (2004); [arXiv:hep-th/0402075].
\end{thebibliography}
\end{document}